\documentstyle[axodraw]{frank}

\newenvironment{comment}[1]{}{}
\def\beq{\begin{equation}}
\def\eeq{\end{equation}}
\def\bea{\begin{eqnarray}}
\def\eea{\end{eqnarray}}
\def\barr{\begin{array}}
\def\earr{\end{array}}
\def\lsim{\mathrel{\vcenter{\hbox{$<$}\nointerlineskip\hbox{$\sim$}}}}

\newcommand{\sm}{Standard Model}
\newcommand{\cm}{center of mass}
\newcommand{\xs}{cross section}
\newcommand{\lc}{linear collider}
\newcommand{\lk}{leptoquark}
\newcommand{\pe}{\mbox{$e^+e^-$}}
\newcommand{\ee}{\mbox{$e^-e^-$}}
\newcommand{\ep}{\mbox{$e^-\gamma$}}
\newcommand{\pp}{\mbox{$\gamma\gamma$}}

\begin{document}

\begin{flushright}
PSI-PR-96-29\\
IFP-721-UNC\\
WUE-ITP-96-024\\
October 1996
\end{flushright}

\vfill

\begin{frontmatter}
\title{Production of Two Non-Conjugate Leptoquarks 
	in $e^-e^-$ Collisions}
\author{Frank Cuypers}
\address{{\tt cuypers@pss058.psi.ch}\\
        Paul Scherrer Institute,
        CH-5232 Villigen PSI,
        Switzerland}
\author{Paul H. Frampton}
\address{{\tt frampton@physics.unc.edu}\\
        Department of Physics and Astronomy, University of North Carolina,\\ 
	Chapel Hill, NC, 27599--3255, USA}
\author{Reinhold  R\"uckl}
\address{{\tt rueckl@physik.uni-wuerzburg.de}\\
        Institut f\"ur Theoretische Physik, Universit\"at W\"urzburg,\\ 
	D-97074 W\"urzburg, Germany}
\begin{abstract}
We study the production of scalar and vector
leptoquarks in $e^-e^-$ scattering.
We use the most general  couplings to the known fermions
which are dimensionless, baryon and lepton number
conserving, and $SU(3)_c \otimes SU(2)_L \otimes U(1)_Y$
invariant.  Expressions are presented for  production   cross
sections and predictions are given for future linear colliders.
\end{abstract}
\end{frontmatter}

\vfill
\clearpage

\section{Introduction}
\label{sect1}
Leptoquarks are contained in many extensions of the Standard Model which
try to unify lepton and quark degrees of freedom~\cite{LQ1}.
In some of these scenarios the leptoquark states emerge at mass scales
below 1 TeV, a range which can be probed with present and future high
energy colliders. Since none of the possible extensions of the
Standard Model are selected yet
by experimental data, the
search for leptoquarks should be carried out in a widely model
independent framework~\cite{BRW}. 
Demanding the \lk\ couplings
to be dimensionless, 
$SU(3)_c \otimes SU(2)_L \otimes U(1)_Y$ invariant, and
baryon ($B$) and lepton ($L$) number conserving,
there are nine scalar and nine vector states
which qualify. 	 
Only a subset of these
is contained in most of the proposed models\footnote{Note that
there are models
which contain {\it all} these states~\cite{FRA2}.}.

There are some very stringent low energy limits 
on the ratios
$m_{LQ}^2/\lambda^2$
of the \lk\ masses
to their couplings to fermion~\cite{sacha}.
However,
these bounds vary greatly 
depending on the structure
of the \lk-fermion coupling matrix.
For some choices 
there are even no limits at all.

In the above framework, the decay pattern of leptoquarks
does not allow a clear distinction of 
the different species in general~\cite{BRW}. 
However, the possible production
channels are rather selective with respect to the different quantum
numbers of the leptoquark states.
More specifically,
the discrimination between the different types of \lk s
will be greatly facilitated
at a \lc\ of the next generation,
when it is
run in its four different modes:
\pe~\cite{BR}, \ep~\cite{EP}, \pp~\cite{PP} and \ee~\cite{EE}.

In this Letter 
we focus on the \ee collider mode \cite{e-e-}
and show that the pair production
of leptoquarks in $e^- e^-$ scattering
is only possible for a
particular subset of combinations complementary to other
leptoquark production reaction.
After
classifying
 the different combinations of leptoquarks
which can be
produced in $e^- e^-$ scattering in Section~2, the production cross
sections  and the discovery potential are
derived in Section~3, and the Conclusion is stated in Section~4.

\section{Classification of Processes}
\label{sect2}
We consider
the effective Lagrangian~\cite{BRW}
\begin{equation}
\label{lag}
{\cal L} = {\cal L}_{|F| = 2} + {\cal L}_{F = 0}  
~,
\end{equation}
which describes the interaction of scalar and vector leptoquarks with
the fermions of the Standard Model referring
to the two possible
classes of \lk s distinguished 
by their fermion number
$F=L+3B$. 
Assuming the conservation laws enumerated in the Introduction
and non-derivative
couplings one has
\begin{eqnarray}
\label{eq2}
{\cal L}_{|F|=2}  &=& ( g_{1L} \bar{q}^c i
\tau_2 l
                +    g_{1R} \bar{u}^c e) S_1
\nonumber\\
               &+&   \tilde{g}_{1R} \bar{d}^c e
\tilde{S}_1
                +    g_{3L} \bar{q}^c i \tau_2
                      \vec{\tau} l \vec{S}_3
\nonumber\\
               &+& ( g_{2L} \bar{d}^c \gamma^{\mu} l
                +    g_{2R} \bar{q}^c \gamma^{\mu} e)
V_{2 \mu}
\nonumber\\
               &+&   \tilde{g}_{2L} \bar{u}^c \gamma^{\mu}
l
                     \tilde{V}_{2 \mu}
\nonumber\\
&+& h.c.,
\end{eqnarray}
\begin{eqnarray}
\label{eq3}
{\cal L}_{F=0}  &=& ( h_{2L} \bar{u} l
                +    h_{2R} \bar{q} i \tau_2 e) R_2
                +    \tilde{h}_{2L} \bar{d} l \tilde{R}_2
\nonumber\\
               &+& ( h_{1L} \bar{q} \gamma^{\mu} l
                +    h_{1R} \bar{d} \gamma^{\mu} e )
U_{1 \mu}
\nonumber\\
               &+&   \tilde{h}_{1R} \bar{u} \gamma^{\mu}
e
                     \tilde{U}_{1 \mu}
                +    h_{3L} \bar{q} \vec{\tau}
                     \gamma^{\mu} l \vec{U}_{3 \mu}  
\nonumber\\
&+& h.c.
\end{eqnarray}
where $f^c = C \overline{f}^T$ and $\tau_i$ are the Pauli matrices.
We use the notation  
$l = ( e_L, \nu_L)$ and $q = ( d_L, u_L)$ 
for the left-handed $SU(2)_L$ lepton and quark doublets, 
and 
$e = e_R$, $d = d_R$ and $u = u_R$ 
for the right-handed singlets.
The family indices are suppressed,
but it should be borne in mind
that in general the constants $g$ and $h$
denote $3 \times 3$ coupling matrices.
The subscript of the scalar \lk\ fields $S$, $\widetilde{S}$, 
$R$, and $\widetilde{R}$, and of the vector fields
$V_\mu$, $\widetilde{V}_\mu$, $U_\mu$, and $\widetilde{U}_\mu$
indicates their $SU(2)_L$ 
singlet, doublet or triplet nature.
The quantum numbers of the 18 individual states in Eqs~(\ref{eq2},\ref{eq3})
have been summarized in Refs~\cite{BRW,BR}.

In $e^-e^-$ scattering, the process of leptoquark pair production is
characterized by the following quantum numbers defined by
the possible initial states:

\begin{eqnarray}
&& Q = -2 \nonumber\\
&& F = ~~2 \nonumber\\
&& (T,T_3) = \left \{ \begin{array}{rr} (0,0) & RR\\
(\frac{1}{2},- \frac{1}{2}) & LR \\ (1,-1)& LL \end{array} \right.
\label{symm}
\end{eqnarray}

with $Q$ the electric charge,
$F = L + 3B$ the
fermion number, $T$ the weak isospin
and $T_3$ its third component.
Here, $R$ and $L$ label the different helicity states of the electrons.
Thus, 
if either only $|F| = 2$ or $F = 0$ leptoquarks exist
they cannot be produced in $e^-e^-$
scattering by $2 \rightarrow 2$ processes.
Pair-production can only take place 
in the presence
of two {\em non-conjugate} \lk s.
The possible leptoquark combinations 
carrying the quantum numbers (\ref{symm}) of a given $e^-e^-$ initial
state are
summarized in Table~1. 

\vspace{4mm}
\begin{center}
\begin{tabular}{||c|c||c|c|c|c||}
\hline \hline
\multicolumn{2}{||c||}{              } &
\multicolumn{4}{c||}{$Q = -1/3 \qquad F = 2$} \\
\cline{2-6}
\multicolumn{1}{||c|}{              } &
\multicolumn{1}{c||}{$
\begin{array}{lr} T &  \\ &  T_3  \end{array}$
}
 &
\multicolumn{1}{c|}{$S_1^{-1/3}$} &
\multicolumn{1}{c|}{$S_3^{-1/3}$} &
\multicolumn{1}{c|}{$V_2^{-1/3}$} &
\multicolumn{1}{c||}{$\widetilde{V}_2^{-1/3}$} \\
\hline \hline
 &
$R_2^{-5/3}$ 
& $\begin{array}{lr} \frac{1}{2} & \\ & -\frac{1}{2} \end{array}$
& $\begin{array}{lr} \frac{1}{2} & \\ & -\frac{1}{2} \end{array}$
& $\begin{array}{lr} 0~          & \\ & ~0           \end{array}$
& $\begin{array}{lr} 1           & \\ & -1           \end{array}$ \\
\cline{2-6} 
$\begin{array}{c} Q = -5/3  \\ F = 0  \end{array}$
& $\widetilde{U}_1^{-5/3}$
& $\begin{array}{lr} 0~          & \\ & ~0           \end{array}$
& $\begin{array}{lr}             & \\ &              \end{array}$
& $\begin{array}{lr}             & \\ &              \end{array}$
& $\begin{array}{lr} \frac{1}{2} & \\ & -\frac{1}{2} \end{array}$ \\
\cline{2-6}
&
$U_3^{-5/3}$
& $\begin{array}{lr} 1           & \\ & -1           \end{array}$
& $\begin{array}{lr} 1           & \\ & -1           \end{array}$
& $\begin{array}{lr} \frac{1}{2} & \\ & -\frac{1}{2} \end{array}$
& $\begin{array}{lr}             & \\ &              \end{array}$\\
\hline \hline
\end{tabular}

\vspace{2mm}
\begin{tabular}{||c|c||c|c|c|c||}
\hline \hline
\multicolumn{2}{||c||}{              } &
\multicolumn{4}{c||}{$Q = -2/3 \qquad F = 0$} \\
\cline{2-6}
\multicolumn{1}{||c|}{              } &
\multicolumn{1}{c||}{$
\begin{array}{lr} T &  \\ &  T_3  \end{array}$
}
 &
\multicolumn{1}{c|}{$R_2^{-2/3}$} &
\multicolumn{1}{c|}{$\widetilde{R}^{-2/3}_2$} &
\multicolumn{1}{c|}{$U_1^{-2/3}$} &
\multicolumn{1}{c||}{$U^{-2/3}_3$} \\
\hline \hline
 &
$\widetilde{S}_1^{-4/3}$ 
& $\begin{array}{lr}             & \\ &              \end{array}$
& $\begin{array}{lr} \frac{1}{2} & \\ & -\frac{1}{2} \end{array}$
& $\begin{array}{lr} 0~          & \\ & ~0           \end{array}$
& $\begin{array}{lr}             & \\ &              \end{array}$ \\
\cline{2-6} 
$\begin{array}{c} Q = -4/3 \\ F = 2 \end{array}$
&
$S_3^{-4/3}$
& $\begin{array}{lr} \frac{1}{2} & \\ & -\frac{1}{2} \end{array}$
& $\begin{array}{lr}             & \\ &              \end{array}$
& $\begin{array}{lr} 1           & \\ & -1           \end{array}$
& $\begin{array}{lr} 1           & \\ & -1           \end{array}$ \\
\cline{2-6}
&
$V_2^{-4/3}$
& $\begin{array}{lr} 0~          & \\ & ~0           \end{array}$
& $\begin{array}{lr} 1           & \\ & -1           \end{array}$
& $\begin{array}{lr} \frac{1}{2} & \\ & -\frac{1}{2} \end{array}$
& $\begin{array}{lr} \frac{1}{2} & \\ & -\frac{1}{2} \end{array}$ \\
\hline \hline
\end{tabular}

\vspace{4mm}
\small
Table~1: Leptoquark pairs obeying Eqs~(\ref{symm})
\normalsize
\end{center}

\section{Production Cross Sections}
\label{sec3}

\begin{figure}[htb]
{\unitlength.5mm\SetScale{1.4185}

{\bf1:}
\begin{picture}(100,40)(-30,16)
\Text(-4,32)[rc]{$e^-_{R}$}
\Text(-4,00)[rc]{$e^-_{R}$}
\ArrowLine(00,32)(24,32)
\ArrowLine(00,00)(24,00)
\ArrowLine(24,32)(24,00)
\Text(28,16)[lc]{$u$}
\DashLine(24,32)(48,32){2}
\Photon(24,00)(48,00){-1}{4}
\Text(52,32)[lc]{$R_2^{-5/3}$}
\Text(52,00)[lc]{$V_2^{-1/3}$}
\end{picture}
\qquad + \qquad crossed

\bigskip\bigskip

{\bf2:}
\begin{picture}(100,40)(-30,16)
\Text(-4,32)[rc]{$e^-_L$}
\Text(-4,00)[rc]{$e^-_R$}
\ArrowLine(00,32)(24,32)
\ArrowLine(00,00)(24,00)
\ArrowLine(24,32)(24,00)
\Text(28,16)[lc]{$u$}
\DashLine(24,32)(48,32){2}
\DashLine(24,00)(48,00){2}
\Text(52,32)[lc]{$R_2^{-5/3}$}
\Text(52,00)[lc]{$S_1^{-1/3}$}
\end{picture}
\qquad
\begin{picture}(100,40)(-30,16)
\Text(-4,32)[rc]{$e^-_L$}
\Text(-4,00)[rc]{$e^-_R$}
\ArrowLine(00,32)(24,32)
\ArrowLine(00,00)(24,00)
\ArrowLine(24,00)(24,32)
\Text(28,16)[lc]{$u$}
\DashLine(24,32)(48,32){2}
\DashLine(24,00)(48,00){2}
\Text(52,00)[lc]{$R_2^{-5/3}$}
\Text(52,32)[lc]{$S_1^{-1/3}$}
\end{picture}

\bigskip\bigskip

{\bf3:}
\begin{picture}(100,40)(-30,16)
\Text(-4,32)[rc]{$e^-_L$}
\Text(-4,00)[rc]{$e^-_R$}
\ArrowLine(00,32)(24,32)
\ArrowLine(00,00)(24,00)
\ArrowLine(24,32)(24,00)
\Text(28,16)[lc]{$d$}
\DashLine(24,32)(48,32){2}
\DashLine(24,00)(48,00){2}
\Text(52,32)[lc]{$\tilde R_2^{-2/3}$}
\Text(52,00)[lc]{$\tilde S_1^{-4/3}$}
\end{picture}

\bigskip\bigskip

{\bf4:}
\begin{picture}(100,40)(-30,16)
\Text(-4,32)[rc]{$e^-_L$}
\Text(-4,00)[rc]{$e^-_R$}
\ArrowLine(00,32)(24,32)
\ArrowLine(00,00)(24,00)
\ArrowLine(24,32)(24,00)
\Text(28,16)[lc]{$d$}
\Photon(24,32)(48,32){1}{4}
\Photon(24,00)(48,00){-1}{4}
\Text(52,32)[lc]{$U_1^{-2/3}$}
\Text(52,00)[lc]{$V_2^{-4/3}$}
\end{picture}
\qquad
\begin{picture}(100,40)(-30,16)
\Text(-4,32)[rc]{$e^-_L$}
\Text(-4,00)[rc]{$e^-_R$}
\ArrowLine(00,32)(24,32)
\ArrowLine(00,00)(24,00)
\ArrowLine(24,00)(24,32)
\Text(28,16)[lc]{$d$}
\Photon(24,32)(48,32){1}{4}
\Photon(24,00)(48,00){-1}{4}
\Text(52,00)[lc]{$U_1^{-2/3}$}
\Text(52,32)[lc]{$V_2^{-4/3}$}
\end{picture}

\bigskip\bigskip

{\bf5:}
\begin{picture}(100,40)(-30,16)
\Text(-4,32)[rc]{$e^-_L$}
\Text(-4,00)[rc]{$e^-_R$}
\ArrowLine(00,32)(24,32)
\ArrowLine(00,00)(24,00)
\ArrowLine(24,32)(24,00)
\Text(28,16)[lc]{$u$}
\Photon(24,32)(48,32){1}{4}
\Photon(24,00)(48,00){-1}{4}
\Text(52,32)[lc]{$U_3^{-5/3}$}
\Text(52,00)[lc]{$V_2^{-1/3}$}
\end{picture}

\bigskip\bigskip

{\bf5':}
\begin{picture}(100,40)(-30,16)
\Text(-4,32)[rc]{$e^-_L$}
\Text(-4,00)[rc]{$e^-_R$}
\ArrowLine(00,32)(24,32)
\ArrowLine(00,00)(24,00)
\ArrowLine(24,32)(24,00)
\Text(28,16)[lc]{$u$}
\Photon(24,32)(48,32){1}{4}
\Photon(24,00)(48,00){-1}{4}
\Text(52,32)[lc]{$U_3^{-5/3}$}
\Text(52,00)[lc]{$V_2^{-1/3}$}
\end{picture}
\qquad
\begin{picture}(100,40)(-30,16)
\Text(-4,32)[rc]{$e^-_L$}
\Text(-4,00)[rc]{$e^-_R$}
\ArrowLine(00,32)(16,16)
\ArrowLine(00,00)(16,16)
\Photon(16,16)(32,16){1}{3}
\Text(26,18)[bc]{$B^{--}$}
\Photon(32,16)(48,32){1}{4}
\Photon(32,16)(48,00){-1}{4}
\Text(52,32)[lc]{$U_3^{-5/3}$}
\Text(52,00)[lc]{$V_2^{-1/3}$}
\end{picture}

\bigskip\bigskip
\bigskip\bigskip

}
\caption{Typical Feynman diagrams 
  for \lk\ pair production in \ee\ collisions.
  The quark generation indices are suppressed.}
\label{feyn}
\end{figure}
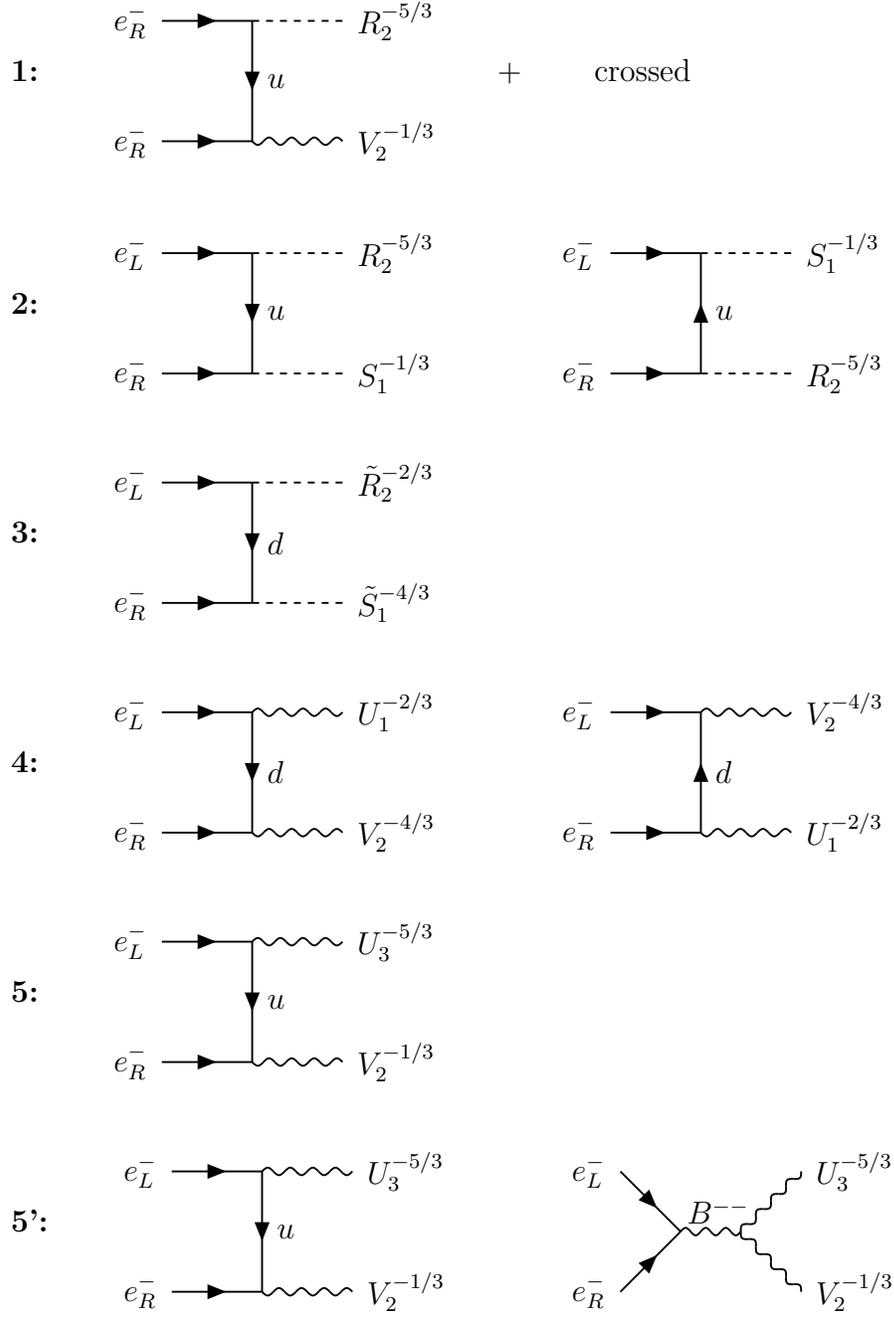

According to Table~1,
there are in total 18 different reactions 
which can take place.
These can be subdivided into five classes 
whose typical Feynman diagrams are shown in
Fig.~\ref{feyn}.
They are uniquely characterizable by the weak isospin $(T,T_3)$ of the 
\lk ~pair, the \lk ~spins $s,$ and the number $n$ of scattering channels:

\begin{eqnarray}
\label{class}
{\bf1 :} \quad && (T,T_3) = (0,0) \quad \mbox{ and } \quad (1,-1)~; 
\nonumber\\
{\bf2 :} \quad && (T,T_3) = (\frac{1}{2},-\frac{1}{2}) \quad s=0 \quad n=2~;  
\nonumber\\
{\bf3 :} \quad && (T,T_3) = (\frac{1}{2},-\frac{1}{2}) \quad s=0 \quad n=1~;
\\
{\bf4 :} \quad && (T,T_3) = (\frac{1}{2},-\frac{1}{2}) \quad s=1 \quad n=2~;
\nonumber\\
{\bf5 :} \quad && (T,T_3) = (\frac{1}{2},-\frac{1}{2}) \quad s=1 \quad n=1~.
\nonumber
\end{eqnarray}

The corresponding integrated \xs s are given by

\begin{equation}
\label{r1}
\sigma_1 = 4G
\left[~
S + L\left( D + 2m_S^2 + 2{m_S^4 \over D} \right)
~\right],
\end{equation}
\begin{equation}
\label{r2}
\sigma_2 = 2G
\left[~
3S + L\left( D + 2{m_0^2m_2^2 \over D} \right)
~\right],
\end{equation} 
\begin{equation}
\label{r3}
\sigma_3 = G
\left[~
2S + LD
~\right],
\end{equation}
\begin{equation}
\label{r4}
\sigma_4 = 8G
\left[~
2S + L\left( D + 2(m_0^2+m_2^2) + 2{(m_0^2+m_2^2)^2 \over D} \right)
~\right], 
\end{equation}
\begin{equation}
\label{r5}
\sigma_5 = G
\left[~
{S\over6} \left( {D^2 \over m_0^2m_2^2} + 12({D\over m_0^2}+{D\over m_2^2})
+ 12({m_0^2\over m_2^2}+{m_2^2\over m_0^2}) - 28 \right)
 - 4LD
~\right].
\end{equation}

Here,

\begin{eqnarray}
\label{par}
G &=& {3\pi\alpha^2\over s^2} \left({\lambda\over e}\right)^4 
\nonumber\\
D &=& s - m_0^2 - m_2^2 
\nonumber\\
S &=& \sqrt{D^2-4m_0^2m_2^2} 
\\
L &=& \ln{D+S \over D-S},
\nonumber
\end{eqnarray}

$\lambda = \sqrt{hg}$ being the geometric mean of the
leptoquark-lepton-quark couplings, and
$m_0$, $m_2$, $m_S$ denoting the mass of the $F=0$, $F=2$ and the scalar
leptoquark, respectively. 
Furthermore, $\sqrt{s}$ is the \cm\ energy.

In general, the dependence of the cross sections on the Yukawa couplings
is slightly more involved than what is shown above, where we have dropped
the chirality labels $L$ and $R$. They can
be easily restored from the
Feynman diagrams of Fig.~\ref{feyn} using Eqs~\ref{eq2} and \ref{eq3},
for the processes of 
class 1 ($h_R g_R$ or $h_L g_L$), and 
class 3 and 5 ($h_L g_R$ or $h_R g_L$). 
Class 2 and 4, however, simultaneously involve
two diagrams with different couplings, $h_L g_R$ and $h_R g_L$.
The results for $\sigma_2$ and $\sigma_4$ given
in Eqs~\ref{r2} and \ref{r4}, respectively, apply only to the case 
where $h_L g_R = h_R g_L$.
Obviously, an even clearer discriminating
production pattern arises if the $L$- and $R$-couplings cannot be
sizable simultaneously, 
as suggested by low-energy bounds \cite{sacha}.
In the limit where either $h_L g_R=0$ or $h_R g_L=0$
the \xs s for class 2 and 4 
become those of class 3 and 5 respectively:
$\sigma_2 \to \sigma_3$ 
and
$\sigma_4 \to \sigma_5$.

\begin{figure}[htb]
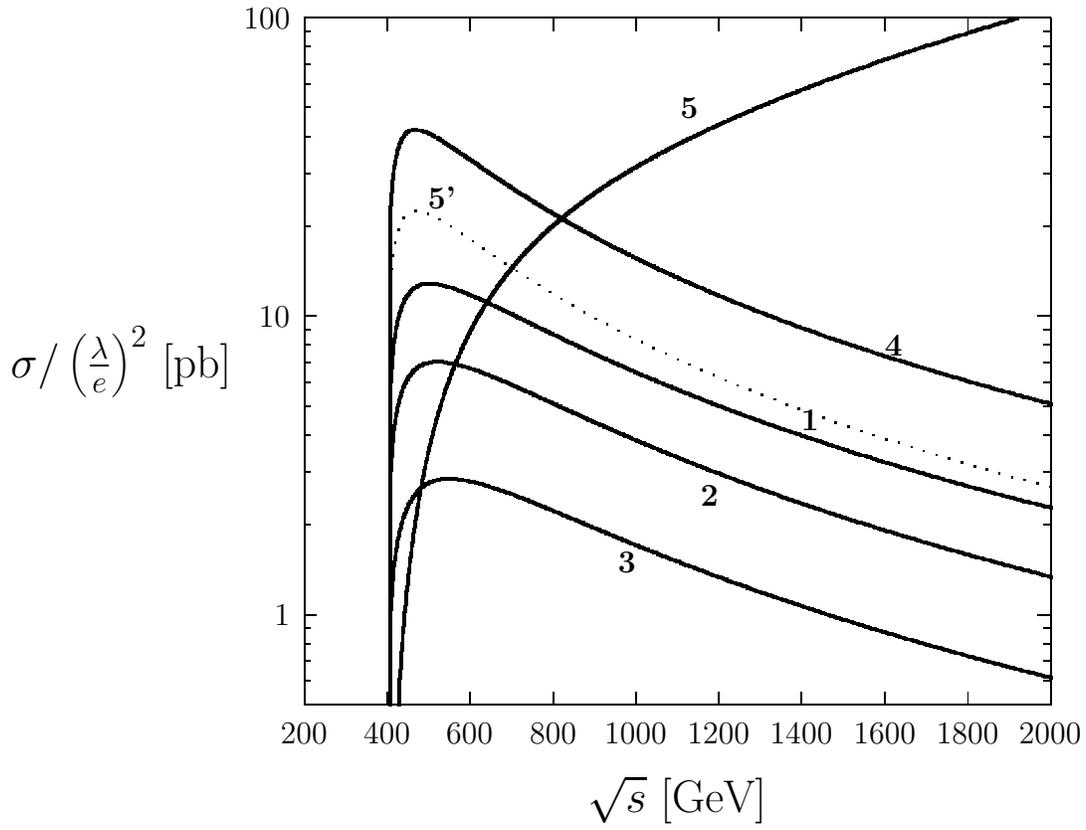

\centerline{\input en.tex}
\caption{Cross section as a function of the collider energy.
  The common mass of the produced leptoquarks is 200 GeV.}
\label{eny}
\end{figure}

\begin{figure}[htb]
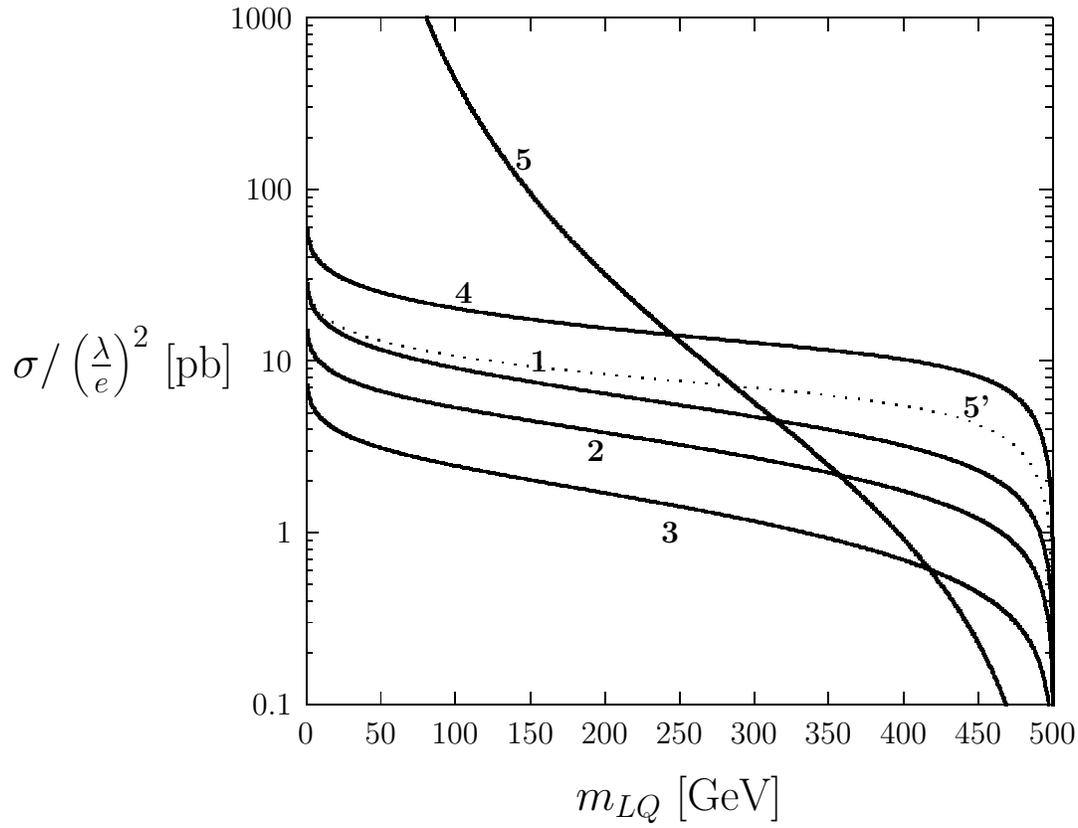

\centerline{\input ma.tex}
\caption{Cross section as a function of the common mass 
  of the produced leptoquarks.
  The collider energy is 1 TeV.}
\label{mass}
\end{figure}

In Figs~\ref{eny} and \ref{mass},
we have plotted the energy and mass dependence
of these cross sections
assuming 100\%\ polarized beams,
equal masses $m_0=m_2=m_S=m_{LQ}$ for both leptoquarks,
and $\lambda=e$.
The electron and quark masses 
are set to zero.
The high energy ($\sqrt{s} \gg m_0,~m_2$) behaviour of these cross
sections is as follows: 

\beq
\label{ash}
\sigma_{1-4} \propto {1\over s}\ln{s\over m_0m_2}
\qquad\qquad\qquad\quad
\sigma_5 \propto s. 
\eeq

At threshold ($\sqrt{s} \approx m_0 + m_2$), on the other hand,
one has

\beq
\label{asl}
\sigma_{1-4} \propto \sqrt{s-(m_0+m_2)^2}
\qquad\qquad
\sigma_5 \propto s-(m_0+m_2)^2 
\eeq

The pathological breaking of perturbative unitarity 
in the processes of class 5
is due to its purely $t$-channel nature.
For non-gauge \lk s,
this behaviour is not worrisome,
since in this case new physics effects
are anyway expected to set in 
at some higher energy scale.

For gauge \lk s, though,
a gauge bilepton~\cite{FRA2} 
may be exchanged in the $s$-channel.
If it couples 
with strength $\lambda$ to the leptoquarks and electrons,
the complete cross section becomes:

\begin{eqnarray}
\sigma_{5'} = G
& \Biggl\{ &
{S\over6} 
\biggl[ 
  62 
+ {2m_B^2 - m_0^2 \over m_2^2}
+ {2m_B^2 - m_2^2 \over m_0^2}
- {m_B^4 \over m_0^2m_2^2}
\nonumber\\
&& \quad + {2\over s-m_B^2} \Bigl(
  5m_0^2 + 5m_2^2 + 22m_B^2
- {m_B^6 \over m_0^2m_2^2}
\nonumber\\
&& \quad\qquad\qquad 
+ {9m_0^2m_B^2 - 5m_0^4 - 3m_B^4 \over m_2^2}
+ {9m_2^2m_B^2 - 5m_2^4 - 3m_B^4 \over m_2^2}
\Bigr)
\nonumber\\
&& \quad - {1\over (s-m_B^2)^2} \Bigl(
  8m_0^4 + 8m_2^4 - 32m_B^4 - 32m_0^2m_B^2 - 32m_2^2m_B^2
\nonumber\\
&& \quad\quad\quad\quad\quad
- 18m_0^2m_2^2 + {m_B^8 \over m_0^2m_2^2}
\nonumber\\
&& \quad\qquad\qquad 
+ {8m_0^4m_B^2 - 18m_0^2m_B^4 + m_0^6 + 8m_B^6 \over m_2^2}
\nonumber\\
&& \quad\quad\quad\quad\quad
 + {8m_2^4m_B^2 - 18m_2^2m_B^4 + m_2^6 + 8m_B^6 \over m_0^2}
\Bigr)
\biggl] 
\nonumber\\
& + & 
4L \left[ D+2(m_0^2+m_2^2)+2{m_0^2m_2^2 + m_2^2m_B^2 + m_B^2m_0^2 \over s-m_B^2} \right]
\quad\Biggr\} 
~.
\end{eqnarray}

This \xs\ has the same high energy and threshold behaviours
as the class 1 -- 4 reactions.
It is also shown on the plots
as the dotted curve,
for the bilepton mass $m_B=m_0=m_2$.

The leptoquarks which can be produced according to the diagrams
of Fig.~\ref{feyn} all 
decay into a charged lepton and a jet
with a substantial branching ratio.
If they couple only to the first generation of leptons,
the decay lepton is an electron.
Close to threshold,
the signature is thus
a pair of high transverse momentum electrons,
a pair of high transverse momentum jets
and no missing energy.
If we cut out the (very few) jet pairs with an invariant mass
around the $Z^0$,
the lowest order remaining background 
originates from the
$e^-e^- \to e^-e^-q\bar q$
continuum.
It should be tiny.
If one allows for a leptoquark-fermion coupling matrix
which mixes different families,
like a diagonal matrix for instance,
the reaction does not necessarily conserve lepton flavour.
In this case there is no \sm\ background at all.

\begin{figure}[htb]
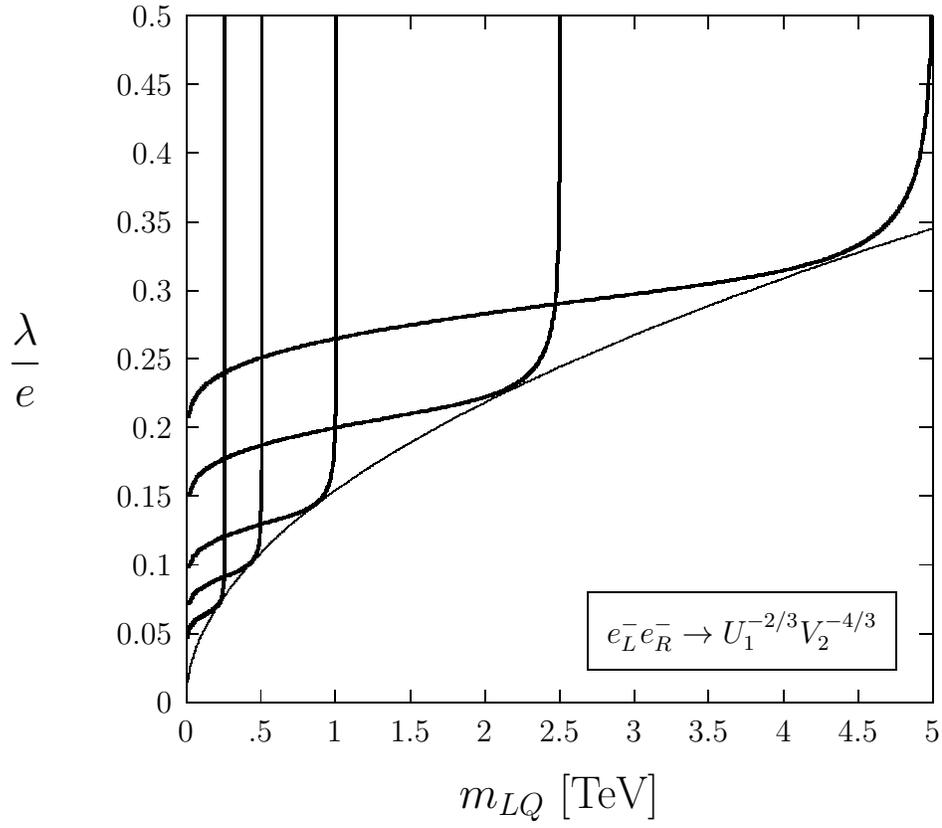

\centerline{\input lim.tex}
\caption{Loci of $\sigma_4=1$ fb,
  as a function of the common leptoquark mass and coupling to fermions.
  The collider energies are .5, 1, 2, 5 and 10 TeV.
  The thinner osculating parabola is given by Eq.~(\protect\ref{osc}).}
\label{lim}
\end{figure}

To estimate the discovery potential,
we have plotted in Fig.~\ref{lim}
the boundary in the $(m_{LQ},\lambda/e)$ plane, 
above which 
more than 10 events 
are produced in the class 4 reaction.
For this we consider 5 different collider energies
and assume 10 fb$^{-1}$ of accumulated luminosity.
Again the leptoquarks have the same common mass.
In general,
an osculating parabola can be fitted to this family of curves.
Its equation is approximately given by

\begin{equation}
\label{osc}
{\lambda \over e}  =  0.35  \sqrt{m_{LQ}/\mbox{TeV}} 
\left(N \over A~{\cal L}/\mbox{fb}^{-1}\right)^{1/4}
~,
\end{equation}

where 
$N$ is the average number of events,
$\cal L$ is the integrated luminosity and
$A=6,~3,~1,~24,~12$
for reactions
belonging to class 1, 2, 3, 4, 5',
respectively.
These numbers originate from the omitted factors
in Eqs~(\ref{ash}).

Eq.~(\ref{osc}) conveniently summarizes 
the \lk\ discovery potential of \ee\ collisions,
provided the common \lk\ mass
lies safely below the pair production threshold 
of the largest attainable \cm\ energy
$m_{LQ}<0.43\sqrt{s}$.
This limit approximately corresponds to the intersection
of the osculating parabola
with each curve of a certain \cm\ energy $\sqrt{s}$.

Assuming five events should suffice to establish a discovery,
an average number of $N=9.15$ Poisson distributed events 
is needed,
such that {\em at least} 5 events
are observed with 95\%\ probability. 
A typical \pe\ design luminosity scaling relation is~\cite{ZDR}

\beq
\label{lumpe}
{\cal L}_{e^+e^-} \mbox{ [fb$^{-1}$]} = 200 s \mbox{ [TeV$^2$]} 
~.
\eeq

Accounting for the luminosity loss
due to anti-pinching at the interaction point,
we have for the \ee\ luminosity~\cite{jim}

\beq
\label{lumee}
{\cal L}_{e^-e^-} = {{\cal L}_{e^+e^-} \over 2}
~.
\eeq

Taking into account the energy scan capability of a \lc\
whose maximum \cm\ energy is $\sqrt{s}$,
non-conjugate \lk s can thus be discovered in \ee\ scattering
with 95\%\ confidence
up to the mass

\beq
\label{limit}
m_{LQ}~
\lsim~
\sqrt{s}~
\min\left(
27 \left({\lambda \over e}\right)^2 \sqrt{A}
~,~
0.43
\right)
~.
\eeq

The first upper bound corresponds to the osculating parabola
(\ref{osc}),
whereas the second bound corresponds to 
the point in the $(m_{LQ},\lambda/e)$-plane where 
the parabola terminates for the given nominal collider energy.

\begin{comment}{
Of course,
if the couplings are very large
even \lk s in the mass range
$0.43\sqrt{s}<m_{LQ}<0.5\sqrt{s}$
can be observed.
Similarly,
if the produced \lk s have very different masses
Eq.~(\ref{limit}) becomes invalid.
}\end{comment}

\section{Conclusion}

Leptoquark production provides another good example for the 
complementarity of the $e^+e^-$ and $e^-e^-$ modes of linear
colliders. Trough lowest-order pair-production, $e^+e^-$ collisions 
mainly probe the interactions of leptoquarks with the electroweak
gauge bosons of the Standard Model, while $e^-e^-$ collisions allow for
powerful tests of the couplings of leptoquarks to electrons and quarks. 

Assuming the $SU(3)_c \otimes SU(2)_L \otimes U(1)_Y$ symmetry of the 
Standard Model to be preserved by the leptoquark interactions,
the strength of the couplings to the gauge bosons is fixed~\cite{BR} by the 
gauge principle, whereas the couplings to fermion fields remain
essentially undetermined. It is therefore important to have independent
tests of the latter. Such tests are possible by analyzing leptoquark
decays and, more directly, by studying the production
in the
$e^-e^-$ and $eq$~\cite{BRW} channels. These channels are particularly
selective with respect to the fermion number and the family structure of the 
leptoquark-fermion interactions. Moreover, if polarized beams are
available they also map out the chiral properties in an
unambiguous way.

In this Letter, this is shown for the $e^-e^-$ mode of future
linear colliders in the TeV-energy range.
Unlike in $e^+e^-$ reactions where leptoquarks are produced in
conjugate pairs, two non-conjugate leptoquarks emerge from
$e^-e^-$ scattering.
In addition to the Standard Model gauge symmetry, 
we have assumed, for this study,  lepton and
baryon number conservation and non-derivative 
Yukawa-type leptoquark-fermion couplings. 

If the couplings are of electromagnetic strength one can expect
cross sections which are sufficiently large to be observed
up to the kinematical pair production threshold.
In this
case one also predicts sizable effects in $e^+e^-$ collision
on top of the production cross sections 
due to the exchange of gauge bosons.
Comparison of measurements in the two modes will then
provide useful cross-checks.      
However, if the Yukawa couplings are considerably weaker than
the electromagnetic coupling, they are not visible in $e^+e^-$
production, but can still be tested in $e^-e^-$ scattering down to 
${\lambda / e} \approx 0.1$. 
Furthermore, if the leptoquarks couple to only
one lepton and quark helicity state, only a fraction of the class 1 -- 5
reactions considered here can proceed revealing clearly the chiral
structure of the leptoquark-fermion interactions.   
 
In summary, although $e^+e^-$ is the better discovery mode, 
a high-energy $e^-e^-$ collider is superior in finding out the strength
and structure of the leptoquark couplings to normal matter.
The latter is important in understanding the elementarity or
compositeness of these hypothetical particles.
              
\vspace{10mm}

The work of P.H.F. was supported in part 
by the U.S. Department of Energy, 
under Grant DE-FG05-85ER-40219, Task B.
The work of R.R. was supported 
by the Bundesministerium f\"ur Bildung und Forschung (BMBF) 
under contract 05 7WZ91P(0).


\clearpage
 
\clearpage

\clearpage

\clearpage

\end{document}